\documentstyle[aps,multicol,amsfonts,graphics,epsfig,amsmath]{revtex}
\newcommand{\fs}{\footnotesize}

\begin{document}
\setcounter{page}{466} SOVIET PHYSICS JETP\hfill VOLUME 30, NUMBER 3\hfill MARCH,
1970\\[5mm] {\sl NONLINEAR INTERFERENCE EFFECTS IN EMISSION, ABSORPTION, AND
GENERATION SPECTRA}\\[3mm] T. Ya. POPOVA, A. K. POPOV, S. G. RAUTIAN, and R. I.
SOKOLOVSKII\\[3mm] \hspace*{15mm}Institute of Semiconductor Physics, Siberian
Division, USSR Academy of Sciences\\[3mm] \hspace*{15mm}Submitted December 27,
1968\\[3mm] \hspace*{15mm}Zh. Eksp. Teor. Fiz. 57, 850-863 (September, 1969)

\begin{abstract}
\noindent Nonlinear effects in emission and absorption spectra of gaseous
systems are considered. It is shown that level splitting can be
detected spectroscopically even if it is below the Doppler width.
Conditions for distinguishing interference effects from those due
to nonequilibrium velocity distribution are determined. In the case
of large Doppler broadening the correction for atomic motion is
equivalent to the substitution of an "effective immobile atom" for
the moving atom ensemble. The spectral manifestation of nonlinear
effects is analyzed in detail. The influence of nonlinear
interference effects on the generation characteristics in the
presence of external field is investigated.
\end{abstract}

\begin{multicols}{2}
\narrowtext
\section{INTRODUCTION}
The changes in the emission and absorption spectra of a gas placed in a strong
electromagnetic field are the result of three effects. One consists of the formation
of a nonequilibrium velocity distribution (Bennett'e "holes" and ``peaks''$^{[1]}$).
This factor significantly influences the spectral characteristics of lasers and was
studied in detail by many authors. The second effect stems from the splitting of
atomic levels; it was directly observed in the optical portion of the spectrum only
very recently$^{[2,3,]}$ in the case of potassium atoms placed in the tremendous
fields of a ruby laser. In gas lasers the fields are weaker, level splitting is much
smaller than the Doppler line width, and the observability of the effect is not a
simple matter. For example, according to Feld and Javan$^{[4]}$, splitting is not
possible at all in this case. This conclusion however is the consequence of an error
in their calculations (see discussion of (3.4) below). Finally, the third effect of a
strong external field consists in the fact that the probability of absorption or
emission of photons turns out to depend not only on level populations but also on the
polarization induced by the external field, i.e., on the nonlinear interference
effect (NIE)$^{[5-7]}$. This effect is the subject of the present paper.

The interest in NIE is due to several causes. First, {\it it is this effect that is
responsible for causing the spectral densities of Einstein coefficients, of
absorption or emission to be different frequency functions leading to characteristic
changes in the pure emission or absorption lines} [7-9]. The NIE contribution should
depend significantly oh the relaxation characteristics$^{[7]}$, providing new
opportunities to study collisions. For gas systems with large Doppler broadening the
theory predicts an angular anisotropy of spectral characteristics and a possibility
of obtaining an extremely sharp structure$^{[4-6,10]}$. Although the early
experiments with spontaneous$^{[4,11,12]}$ and stimulated emission$^{[13]}$ have so
far failed to provide a quantitative verification of the theory, they have
undoubtedly established the existence of the anisotropy effect.

The present work investigates NIE in gaseous systems and considers
the problem under what conditions the plays a major role. It is
shown that under certain conditions the velocity (distribution of
atoms In a strong field does not change at all while the
interference effects remain.

\section{GENERAL EXPRESSIONS}
\begin{figure}
  \centering
\includegraphics[width=30mm]{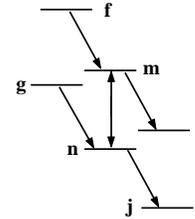}
\caption{Term diagram.}\label{f1}
\end{figure}

We consider the photon emission of two monochromatic fields interacting with an atom
whose term system is shown in Fig. 1. One of the two fields is regarded as strong and
it resonates with the $m-n$ transition, the matrix element of interaction (traveling
wave) is

{\fs
\begin{eqnarray}\label{e2.1}
  V_{mn}\exp\{i\omega_{mn}t\}=-G\exp\{-i(\Omega t-kr)\}.\nonumber\\
  G=d_{mn}E/2\hbar,\quad \Omega=\omega_\mu-\omega_{mn}.
\end{eqnarray}}
We are interested in emission or absorption of photons of a field resonating with one
of the four transitions. $n—j$, $m-l$, $f—m$, and $g—n$ (Fig.\ref{f1}). For example
in the case of $n—j$

{\fs
\begin{eqnarray}\label{e2.2}
  V_{nj}\exp\{i\omega_{nj}t\}=-G_\mu\exp\{-i(\Omega_\mu t-k_\mu r)\},\nonumber\\
  G_\mu=d_{nj}E_\mu/2\hbar,\quad \Omega_\mu=\omega_\mu-\omega_{nj}
\end{eqnarray}}
The system of equations for the density matrix has the form

{\fs
\begin{eqnarray}
&&L_{jj}\rho_{jj}=V_{nj}\rho_{nn}+q_j,\nonumber\\
&&L_{jn}\rho_{jn}=-iV_{mn}\exp\{i\omega_{mn}t\}\rho_{jm}=\nonumber\\
&&=iV_{nj}^*\exp\{-i\omega_{nj}t\}(\rho_{nn}-\rho_{jj}),\nonumber\\
&&L_{jm}\rho_{jm}=-iV_{mn}^*\exp\{-i\omega_{mn}t\}\rho_{jn}=\nonumber\\
&&=-iV_{nj}^*\exp\{-i\omega_{nj}t\}\rho_{nm}; \label{e2.3}
\end{eqnarray}

\begin{eqnarray}
&&L_{mm}\rho_{mm}=+2Re[iV_{mn}\exp\{i\omega_{mn}t\}\rho_{nm}]=q_m,\nonumber\\
&&L_{nn}\rho_{nn}=-2Re[iV_{mn}\exp\{i\omega_{mn}t\}\rho_{nm}]=
q_n+\gamma_{mn}\rho_{mm},\nonumber\\
&&L_{nm}\rho_{nm}=iV_{nm}\exp\{-i\omega_{mn}t\}(\rho_{nn}-\rho_{mm})=q_m,\nonumber\\
&&L_{ik}=\partial/\partial\,t+v\nabla+\Gamma_{ik},\quad
\Gamma_{ll}\equiv\Gamma_l,\label{e2.4}
\end{eqnarray}}
$\Gamma_{ik}$ are transition widths and $q_i$ is the rate of excitation of atoms to
the state $i$, $\rm v$.

According to (\ref{e2.3}) and (\ref{e2.4}) the field $V_{jn}$ does not affect the
population ("weak field"). Therefore the entire system of equations was found to be
split up; eqs. (\ref{e2.4}) include only $\rho_{mm}$, $\rho_{nn}$, and $\rho_{nm}$,
and the solution of the system serves as a "source" for the computation of
$\rho_{jm}$, $\rho_{jn}$ and $\rho_{jj}$ from (\ref{e2.3}). In the case of
(\ref{e2.1}) and (\ref{e2.2}) the system (\ref{e2.3})— (\ref{e2.4}) reduces to
equations whose solution has the form

{\fs
\begin{eqnarray}
&&\rho_{jj}=n_j+\dfrac{\gamma_{nj}}{\Gamma_j}\rho_{nn},\nonumber\\
&&\rho_{nn}=n_n+\dfrac{2\pi G^2}{\Gamma_n\sqrt{1+\ae}}\left(
1-\dfrac{\gamma_{mn}}{\Gamma_n}\right)(n_m-n_n)W_B(v),\nonumber\\
&&\rho_{mm}=n_m-\dfrac{2\pi G^2}{\Gamma_m\sqrt{1+\ae}}(n_m-n_n)W_B(v),\nonumber\\
&&\rho_{nm}=r_{nm}\exp\{-i(\Omega t-k r)\},\nonumber\\
&&r_{nm}=iG(\rho_{mm}-\rho_{nn})/(\Gamma+i\Omega') \label{e2.5}
\end{eqnarray}}
where

{\fs
\begin{eqnarray}
&&W_B(v)=\Gamma_B/\pi[\Gamma_B^2+(\Omega-kv)^2],\
\Gamma_B=\Gamma\sqrt{1+\ae},\nonumber\\
&&\Gamma\equiv\Gamma_{nm},\quad \Omega'=\Omega-kv,\nonumber\\
&&{\Omega_\mu}'=\Omega_\mu-k_\mu v,\
\ae=\tau^2G^2=\frac{2(\Gamma_m+\Gamma_n-\gamma_{mn})}{\Gamma_m\Gamma_n\Gamma},\nonumber\\
&&n_i=\frac{q_i(v)}{\Gamma_i}+\frac{\gamma_{ki}}{\Gamma_i}
\cdot\frac{q_k(v)}{\Gamma_k} \label{e2.6}
\end{eqnarray}}
The quantities $n_i(v)$ represent velocity distributions of
atoms in the absence of a strong field $(G=0)$ determined
by excitation processes $q_i(v)$.

The emission (absorption) power is determined by the general formula
{\fs
\begin{equation}\label{e2.7}
  w_{nj}=-2\hbar\omega_{nj}Re\langle
  iV_{nj}\exp\{i\omega_{nj}t\}\rho_{jn},
  \rangle
\end{equation}}
\noindent where the angle brackets designate averaged velocities $v$ of atoms. Using
the system (\ref{e2.3}) we can express $\rho_{jn}$ in terms of (\ref{e2.5}) and
obtain an expression for power (\ref{e2.7}) in the form
{\fs\begin{equation}\label{e2.8}
  w_{nj}=2\hbar\omega_{nj}|G_\mu|^2Re\left\langle
 \frac{[\Gamma_{jm}+i({\Omega_\mu}'+\Omega')](\rho_{nn}-\rho_{jj})-iGr_{nm}}
 {[\Gamma_{jm}+i({\Omega_\mu}'+\Omega')][\Gamma_{jn}+i{\Omega_\mu}']+G^2}
  \right\rangle.
\end{equation}}
Equation (\ref{e2.8}) clearly reflects the classification of effects due to the
external field. The denominator contains squares $\Omega_\mu$ terms, i.e., it contains
resonances at two frequencies. This can be interpreted as a splitting of the atom
levels in the external field The numerator in (\ref{e2.8}) contains two terms with
significantly different properties. The first term is proportional to the population
difference $\rho_{nn}-\rho_{jj}$ containing Bennett's "holes," as reflected in the
factor $W_B(v)$ (henceforth called the Bennett distribution). The second term
proportional to $r_{nm}$ varies only the the shape but not its integral intensity,
since {\fs $$
\int\limits_{-\infty}^{+\infty}w_{nj}d\Omega_{\mu}=2\pi\hbar\omega_{nj}|G_{\mu}|^2
<\rho_{nn}-\rho_{jj}>.$$} The fact that this term appeared and its property are not
at all specific to the special case under consideration. According to (\ref{e2.3})
the "sources" that "excite" $\rho_{jm}$ and $\rho_{jn}$ are both the population
difference $\rho_{nn}-\rho_{jj}$ and the non-diagonal element $\rho_{nm}$ stimulated
by the strong field for any spectral composition of the strong field. Therefore
$w_{nj}$ contains $\rho_{nm}$ also in the general case, and not only in a
monochromatic field. We can say that this term reflects the "coherence" that is
contributed to the atomic state by the strong field, so that a weak field "mixes" the
m and j states as well as the n and j elates. The last circumstance causes
oscillations at the frequency $\omega+\omega_{\mu}$. The above properties of the term
with $r_{nm}$ allow us to call the associated phenomena nonlinear interference
effects.

We can regard (\ref{e2.8}) as the difference between the
number of acts of emission and absorption of the $\hbar\omega_{\mu}$
photon. All the terms of $w_{nj}$ except $\rho_{jj}$ determine emission
processes. Conversely terms associated with $\rho_{jj}$
control the weak field energy absorption rate. According
 to (\ref{e2.8}) only the level splitting effect stands out in
the absorption probability$^{[2,3,6,14]}$. This is due to the fact that absorption
corresponds to the transition from the unexcited level $j$ to excited level $n$. NIE
is due to the reverse transition from an excited to unexcited state, i.e., in the case
when $n - j$ are contained only in the emission. {\it Therefore the line shapes of
pure emission and absorption turn out to be different due to NIE. The sign of their
difference, i.e., of $w_{nj}$, is determined not only by the sign of population
difference $\rho_{nn}-\rho_{jj}$; in particular the sign of $w_{nj}$ can change with
the change of $\Omega_{\mu}$} $^{[7-9]}$.

Equation (\ref{e2.8}) makes it possible to analyze also
spontaneous emission. For this purpose it is merely
necessary to drop the term $\rho_{jj}$ from (\ref{e2.8}) and replace
$|G_{\mu}|^2$ by a quantity corresponding to the atomic interaction with
zero oscillations of the field$^{[15]}$:
$\gamma_{nj}(8\pi^2)^{-1}\Delta\Omega_{\mu}\Delta O$.
Equations for other transitions are
of the same type and can be obtained from (\ref{e2.8}) by a
simple substitution of indices and signs. For example,
$w_{ml}$ is obtained from the substitutions $m\rightarrow n$, $j\rightarrow l$, and
$\Omega '\rightarrow -\Omega '$.
\section{EMISSION AND ABSORPTION LINE SHAPE IN
TRAVELING MONOCHROMATIC WAVE FIELD} \setcounter{equation}{0} We analyze the role of
nonequilibrium  velocity distribution and nonlinear interference effects. We consider
first two directions of $k_{\mu}$ in detail: along and against {\bf k}. The value of
$w_{nj}$ averaged over $\rm v$ for these two directions is

{\fs
\begin{eqnarray}
\label{e3.1}&&w_{nj}^{\pm}=2\hbar\omega_{nj}|G_{\mu}|^2\frac{\sqrt{\pi}}{k{\bar v}}
\exp\left\{-\frac
{\Omega_{\mu}^2}{(k_{\mu}{\bar v})^2}\right\}\times\\ \nonumber
&&\times\{N_n-N_j+(N_m-N_n)Re[F_{\pm}(\Omega_{\mu})+f_{\pm}(\Omega_{\mu})]\},\\
\label{e3.2}&&F_{\pm}+f_{\pm}=\frac{k_{\mu}}{k}\frac{2G^2}{\sqrt{1+\ae}}\times\nonumber\\
&&\times\frac{\Gamma_n^{-1}
(1-\gamma_{mn}/\Gamma_m)[\Gamma_{\pm}+iz]+[1\pm\sqrt{1+\ae}]/2}{[\Gamma_0+iz][\Gamma_{\pm}
+iz]+G^2},\\
\label{e3.3}&&z=\Omega_{\mu}\mp\Omega k_{\mu}/k, \Gamma_0=\Gamma_{jn}+\Gamma_Bk_{\mu}/k,\nonumber\\
&&\Gamma_{\pm}=\Gamma_{jm}+\Gamma_B(k_{\mu}/k\pm1), \ \Gamma_B=\Gamma\sqrt{1+\ae}.
\end{eqnarray}}
The signs + and - in (\ref{e3.2}) correspond to $k_{\mu}$ directed along and against
k; $f_{\pm}$ and $F_{\pm}$ represent the interference term and a term due to the
nonequilibrium addition to the velocity distribution, respectively. Equation
(\ref{e3.2}) is not applicable if $k_{\mu}<k$ and ${\bf k}_{\mu}\cdot{\bf k}<0$.
Velocity averaging can be performed also in this case. However the obtained
expression can be used to some extent in the analysis only if $\ae$ is small Then
(\ref{e3.2}) is valid if $\Gamma_{-}$ is replaced by
$\Gamma_{jm}k_{\mu}/k+(1-k_{\mu}/k)\Gamma_{jn}$, $G = 0$ and $\ae = 0$ everywhere
(except for the common factor $G^2$), and $[1+\sqrt{1+\ae}]/2$ is replaced by
$k_{\mu}/k$.

A comparison of (\ref{e3.2}) with (\ref{e2.8}) shows that $w_{nj}$ has the same
formal structure as the corresponding expression for the fixed atom whose resonant
frequency is converted with respect to the Bennet distribution maximum and which has
the widths $\Gamma_{\pm}$ and $\Gamma_0$ instead of $\Gamma_{jm}$ and $\Gamma_{jn}$
respectively. The physical meaning of $\Gamma_0$ and $\Gamma_{\pm}$ is as follows.
The perturbation theory distinguishes between step-wise and two-photon processes
whose line shape is determined by the factors $<[\Gamma_{jn}+i(\Omega_{\mu}-{\bf
k}_{\mu} \cdot{\bf v})]^{-1}>$ and $<\{\Gamma_{jm}+i[(\Omega_{\mu}+\Omega)-({\bf
k}_{\mu}+{\bf k})\cdot {\bf v}]\}^{-1}>$. In our case the averaging is carried out
essentially with the Bennett distribution (since $\Gamma_B\ll k_{\mu}\bar v$) and the
result of the averaging is $[\Gamma_0+iz]^{-1}$ and
$[\Gamma_{\pm}+iz]^{-1}$$^{[16]}$. Consequently $\Gamma_0$ is the line width of a
step-wise transition that is the sum of the width $\Gamma_Bk_\mu/k$ of the velocity
distribution converted with respect lo Doppler shifts in the $\omega_{nj}$ region and
the natural width $\Gamma_{jn}$ of the $n - j$ transition. Correspondingly
$\Gamma_\pm$ is the line width of two-photon transition consisting of the natural part
$\Gamma_{jm}$ and the Doppler part $\Gamma_B(k_\mu/k\pm1)$. Thus the physical meaning
of the analogy between (\ref{e3.2}) and the line shape of an "effective atom" is
quite clear. The "effective atom" represents the group of atoms that interact with a
strong field. The "effective atom" has the same system of terms as in Fig.\ref{f1}
except that the widths are changed in accordance with the Bennett distribution and
frequency-correlated properties of the step-wise and two-photon processes$^{[16]}$.

Just as in the case of an individual atom, the step-
wise and two-photon processes in the "effective atom"
cannot be considered independently if $G$ is sufficiently
large$^{[16]}$. In fact the numerator in (\ref{e3.2}) contains $G^2$ and
its expansion in terms of simple fractions

{\fs
\begin{eqnarray}
&&\frac{1}{[\Gamma_0+i z][\Gamma_\pm+iz]+G^2}\\ \nonumber
&&=\frac{1}{(z_1+i z)(z_2+i
z)}\\ \nonumber
&&=\frac{1}{z_1-z_2}\left[\frac{1}{z_2+i z}-\frac{1}{z_1+i
z}\right].\\ \nonumber
&&z_{1,2}=1/2\{\Gamma_0+\Gamma_\pm\pm\sqrt{(\Gamma_0-\Gamma_\pm)^2}-4G^2\}\label{e3.4}
\end{eqnarray}}
yields resonant numerators with $z_1,\: z_2$ rather than with
$\Gamma_0,\: \Gamma_\pm$. Under certain conditions the radical in (\ref{e3.2}) can
turn out to be imaginary, which would correspond to the
splitting of the levels of an effective atom.

Equation (\ref{e3.2}) shows that when $\gamma_{mn}=\Gamma_m$ the effect of velocity
distribution variation is completely eliminated and only the NIE remains. The
physical meaning of this is quite clear. The external field transfers some atoms from
the upper level to the lower; at the same time however the relaxation transition is
reduced by the same quantity since there are no other channels of decay from the
upper level. On the other hand the polarization stimulated by the field at the
transition $m — n$ does not turn to zero (see (\ref{e2.8}), expression for $r_{nm}$
and NIE remains unchanged. The transition $6p^1P_2^0 - 7s^3S_1$ of mercury,
$\lambda=1.529$, at which generation was observed$^{[7]}$ can serve as an example of
a case in which the condition $\gamma_{mn}=\Gamma_m$ is valid.

\begin{figure}
  \centering
\includegraphics[width=40mm]{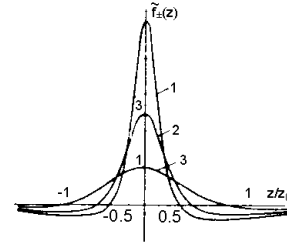}
\caption{Plots of the frequency dependence of the function $\tilde f_\pm$
($z=\Omega_\mu\mp\Omega k_\mu /k$) for real $z_1$ and $z_2$. The curves correspond to the
following values: $1-z_1/z_2=5$; $2-z_1/z_2=2.5$; $3-z_1/z_2=1$.}\label{f2}
\end{figure}

The interference effect. We examine the interference
term $f_\pm(\Omega_\mu)$ in greater detail. Based on (\ref{e3.2}) and (\ref{e3.4})
we have

{\fs
\begin{equation}\label{e3.5}
f_\pm(\Omega_\mu)=\frac{k_\mu}{k}\frac{G^2}{\sqrt{1+\ae}}\frac{1\mp\sqrt{1+\ae}}{z_1-z_2}
\left[\frac{1}{z_2+iz}-\frac{1}{z_1+iz}\right].
\end{equation}}
The line contour of $Re [f_\pm(\Omega_\mu)]$ has the simplest shape when $z_{1,2}$
are real. In this case it follows from (\ref{e3.5}) that the function $Re f_\pm$
changes sign in going from the center of the line to the wings. The sign of $Re f_+$
at the point $z = 0$ is determined by the factor $1+\sqrt{1+\ae}$ and depends
therefore on the relative direction ${\bf k}_\mu$ and ${\bf k}$. When ${\bf
k}_\mu\cdot{\bf k}>0$ the value in the center is negative and in the opposite
direction it is positive. When the values of the external field are small
($\ae\lesssim 1$) we have $Re f_+ \sim \ae^2$ and $Re f_- \sim \ae$.

The function

{\fs $$ f_\pm(z)=\left[\frac{k_\mu}{k}\frac{G^2}{\sqrt{1+\ae}}
\frac{1\mp\sqrt{1+\ae}}{z_1^2}\right]^{-1}Re f_\pm(z) $$} is illustrated in
Fig.\ref{f2} for $z_1/z_2=1;2.5;5$. According to Fig. \ref{f2} the graphs have an
approximately similar shape (the positive maximum in the center and broad negative
wings) for any values of $z_1/z_2$. However the larger $z_1/z_2$ the narrower and
more intense the maximum. When $z_2\ll z_1$ its width is approximately equal to $z_2$
and its intensity in the center is proportional to $z_2^{-1}$. This case seems to be
the most interesting from the practical point of view.

We consider the conditions for which the relation $z_2\ll z_1$
is valid. For the "interference" direction
${\bf k}_\mu\cdot{\bf k}<0$, in which the effect is sharper, the expressions
for $z_{1,2}$ can be represented in the form

{\fs
\begin{eqnarray}\label{e3.6}
&&z_{1,2}=\frac{1}{2}\left\{\Gamma_{jn}+\Gamma_{jm}+\Gamma_B\left(\frac{2k_\mu}{k}-1\right)\right.
\nonumber\\
&&\left.\pm\sqrt{(\Gamma_B+\Gamma_{jn}-\Gamma_{jm})^2-4G^2}\right\}.
\end{eqnarray}}
According to this formula the absence of splitting and
the considerable difference between $z_1$ and $z_2$, are due to
the conditions

{\fs
\begin{equation}\label{e3.7}
\Gamma+\Gamma_{jn}\gg\Gamma_{jm},\: k_\mu\approx k,\: \Gamma^2\ae/G^2=(\Gamma\tau)^2\gg 1.
\end{equation}}
Here the radical in (\ref{e3.6}) can be expanded into a
series:

{\fs
\begin{eqnarray}\label{e3.8}
z_1=\Gamma_{jn}+\Gamma\frac{k_\mu}{k}\sqrt{1+\ae}-\frac{\ae/\tau ^2}{\Gamma
\sqrt{1+\ae}+\Gamma_{jn}-\Gamma_{jm}},\\ \nonumber
z_2=\Gamma_{jm}+\Gamma\left(\frac{k_\mu}{k}-1\right)\sqrt{1+\ae}+\frac{\ae/\tau ^2}{\Gamma
\sqrt{1+\ae}+\Gamma_{jn}-\Gamma_{jm}}.
\end{eqnarray}}
We see from (\ref{e3.8}) that the minimum value of $z_2$ equals the line width of the
forbidden transition $\Gamma_{jm}$. In many cases we can expect that
$\Gamma_{jm}\ll\Gamma{jn}$. Consequently the emission spectrum at the transition $j -
n$ can contain a structure with a considerably smaller width than is typical of the
given transition. The value of $z_2$ increases with the field but much slower than
$z_1$ when $(k_\mu-k)/k\ll 1$.

The amplitude of the interference term

{\fs
\begin{equation}\label{e3.9}
f_-(0)=\frac{k_\mu}{k}\frac{1+\sqrt{1+\ae}}{\sqrt{1+\ae}}\frac{G^2}{z_1z_2}=
\frac{k_\mu}{k}\frac{1+\sqrt{1+\ae}}{\sqrt{1+\ae}}\frac{G^2}{\Gamma_0\Gamma_-+G^2}
\end{equation}}
as a function of $G^2$ is a curve with saturation where one half of the maximum value
is reached approximately for $G^2=\Gamma_0\Gamma_-$. Therefore the ratio
$G^2/\Gamma_0\Gamma_-\equiv \ae_-$ can be interpreted as the saturation parameter of
the effective atom. If $(k_\mu - k)/k\ll 1$ and $\Gamma_0\gg\Gamma_-$, the width
$z_2\approx\Gamma_{jm}[1+\ae_-]$ is also determined by the quantity $\ae_-$. We note
that $\ae_-<\ae$. In tact, according to (\ref{e3.7}) and (\ref{e2.6})

{\fs
\begin{eqnarray}
&&\frac{\ae}{\ae_-}=\Gamma_0\Gamma_-\tau^2=2\left[\Gamma_{jm}+\left(\frac{k_\mu}{k}-1\right)
\Gamma\sqrt{1+\ae}\right]\times\\ \nonumber
&&\left[\Gamma_{jn}+\frac{k_\mu}{k}\Gamma\sqrt{1+\ae}\right]
\frac{\Gamma_m+\Gamma_n-\gamma_{mn}}{\Gamma_m\Gamma\Gamma_n}.\label{e3.10}
\end{eqnarray}}
By virtue of the obvious inequalities $2\Gamma_->\Gamma_m$, $\Gamma_0>\Gamma$ and
$\Gamma_m+\Gamma_n-\gamma_{mn}>\Gamma_n$, the right-hand side in (\ref{e3.10}) is
larger than unity. Therefore as $G^2$ increases the population difference in the
center of the Bennett distribution is equalized first since it is proportional to
$\ae/(1 + \ae)$. The amplitude of the interference term is determined by the ratio
$\ae_-/(1+\ae_-)$, retains its linear dependence up to large values of $G^2$, and
becomes saturated at $\ae_-\approx 1$. At the same time the width of the central
maximum increases, becoming twice as large at $\ae_-=1$ at the same value of the
field.

We now consider the behavior of the interference
term when ${\bf k}_\mu$ is parallel to ${\bf k}$. We first show that $z_1$
and $z_2$ cannot differ significantly in this case. In fact, it
follows from (\ref{e3.4}) that $z_1$ and $z_2$ differ sharply if
$\Gamma_0+\Gamma_+$$\approx\Gamma_0-\Gamma_+$
or $\Gamma_0+\Gamma_+$$\approx\Gamma_+-\Gamma_0$. These conditions in turn
are equivalent to the inequality systems (see (\ref{e3.3}))
$\Gamma_{jm}\gg\Gamma$, $\Gamma_{jm}\gg\Gamma_{jn}$ or $\Gamma_{jn}\gg\Gamma$,
$\Gamma_{jn}\gg\Gamma_{jm}$ which
can be readily shown to be invalid in spontaneous relaxation and in
impact broadening of lines. Consequently
the roots $z_1$ and $z_2$ are of the same order of magnitude
in the direction ${\bf k}_\mu\cdot{\bf k}>0$ and the structure is relatively
not sharp. According to (\ref{e3.5}) the amplitude $f_+(0)$ is

{\fs
\begin{equation}\label{e3.11}
f_+(0)=-\frac{k_\mu}{k}\frac{\sqrt{1+\ae}-1}{\sqrt{1+\ae}}\frac{G^2}{\Gamma_0\Gamma_+ + G^2}.
\end{equation}}
Comparing (\ref{e3.11}) and (\ref{e3.9}) we see that $|f_+(0)|<f_-(0)$,
i.e., the amplitude of the structure in the direction
${\bf k}_\mu\cdot{\bf k}>0$ is always smaller than for ${\bf k}_\mu\cdot{\bf k}<0$.

So far we considered $z_1,\: z_2$ to be real. Now let

\begin{figure}
  \centering
\includegraphics[width=40mm]{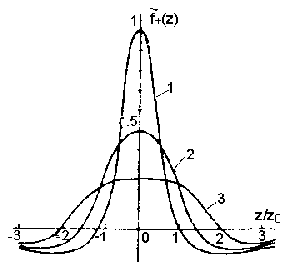}
\caption{Plots of the frequency dependence of the function $\tilde f_\pm$ for
complex $z_1$ and $z_2$ ($z_{1,2}=z_0\pm i\zeta$). The curves correspond to the
following values: $1-\zeta=0$; $2-\zeta =z_0$; $3-\zeta =\sqrt{3}z_0$.}\label{f3}
\includegraphics[width=40mm]{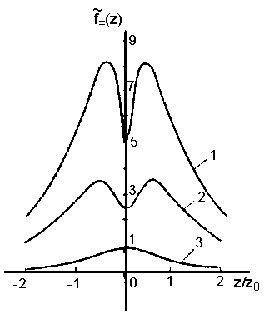}\hfill
\includegraphics[width=40mm]{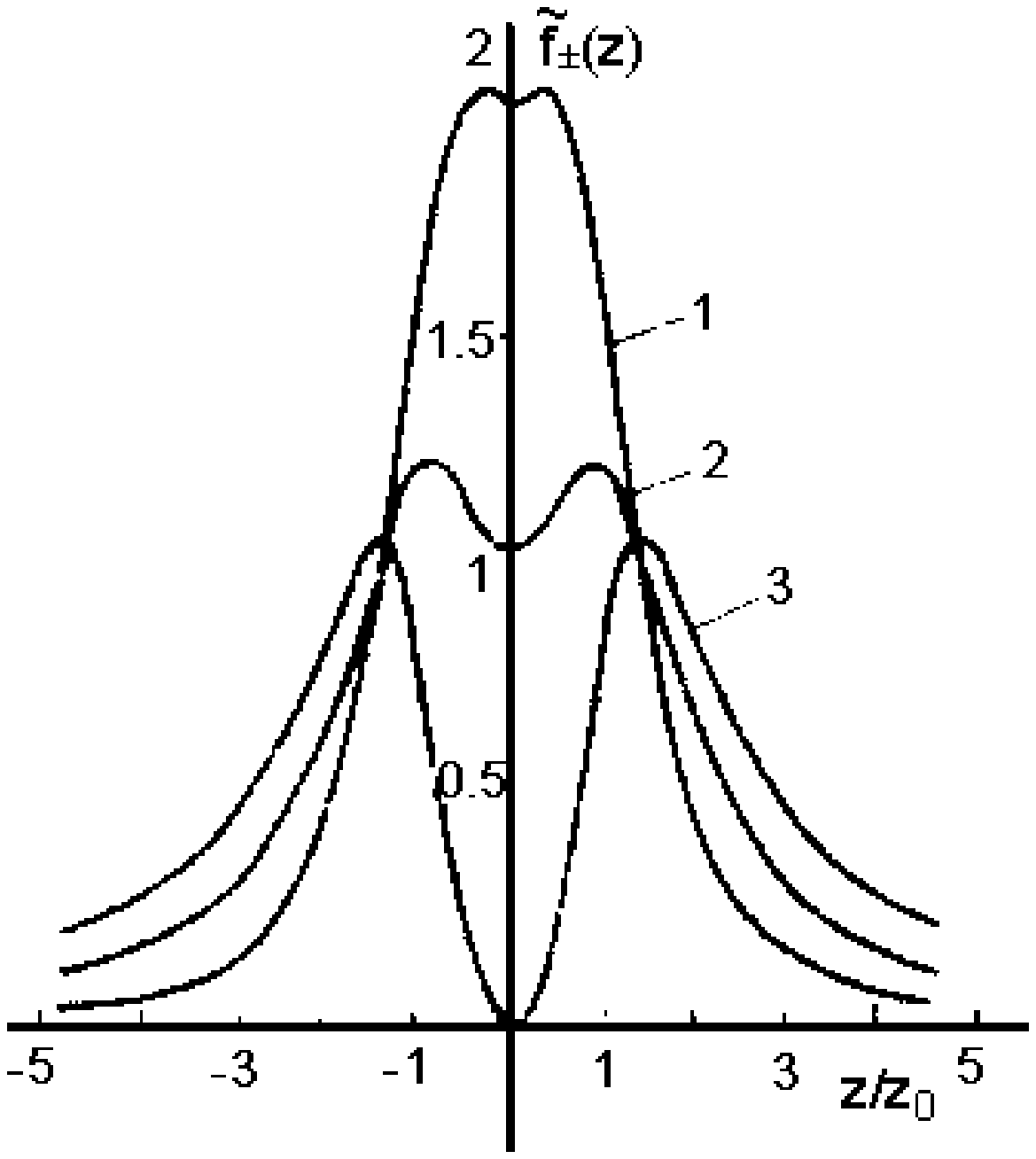}
\label{f4}\caption{Plots of the frequency dependence of the function
$\tilde f_\pm$ for real $z_1$ and $z_2$. The curves correspond to the
following values: $1-z_1/z_2=5$; $2-z_1/z_2=2.5$; $3-z_1/z_2=1$.}
\label{f5}\caption{Plots of the frequency dependence of the function
$\tilde f_\pm$ for complex $z_1$ and $z_2$, $\zeta =z_0$. The curves correspond to the
following values: $1-c=-1$; $2-c=0$; $3-c=1$}
\end{figure}

{\fs
\begin{eqnarray}
\label{e3.12}&&z_{1,2}=z_0+i\zeta, z_0=(\Gamma_0+\Gamma_\pm)/2,\nonumber\\
&&\zeta=\sqrt{G^2-(\Gamma_0-\Gamma_\pm)^2/4},\\
\label{e3.13}&&Re f_\pm(z)=\frac{k_\mu}{k}\frac{1\mp\sqrt{1+\ae}}{\sqrt{1+\ae}}\frac{G^2}
{2\zeta}\times\nonumber\\
&&\times\left[\frac{z+\zeta}{z_0^2+(z+\zeta)^2}-\frac{z-\zeta}{z_0^2+(z-\zeta)^2}\right].
\end{eqnarray}}
The general shape of the graph $Re f_\pm$ depends on the
ratio $\zeta/z_0$, as is apparent from Fig.\ref{f3}. When $\zeta/z_0$ is
small the contours are qualitatively indistinguishable
from the case of real, but similar, $z_1,\: z_2$ (see curves 1
and 2 in Fig.\ref{f3}). It is of interest therefore to determine
the maximum possible values for the ratio $\zeta/z_0$S. We can
show using (\ref{e3.12}) and (\ref{e3.2}) that under the most favorable
conditions $\zeta\leq\sqrt{3}z_0$. The curve in Fig.\ref{f3} corresponding
to $\zeta=\sqrt{3}z_0$ indicates the maximum effect of line splitting.
The "fuzzy" splitting of the interference term has
a physical meaning: the increasing $G^2$ is accompanied
by a rise in the atomic level splitting occurring together, however,
with an increase in the line widths of
effective atom, $\Gamma_0$, and $\Gamma_\pm$ due to the broadening of
Bennett distribution (see (\ref{e2.6})). Nevertheless we can
observe level splitting even with a large Doppler broadening since
the shape of curve 3 in Fig.\ref{f3} is still significantly different
from the others.

Nonequilibrium addition to the velocity distribution.
We turn to the term $F_\pm(\Omega_\mu)$ in (\ref{e3.2}):

{\fs
\begin{eqnarray}\label{e3.14}
&&F_\pm(\Omega_\mu)=\frac{k_\mu}{k}\Gamma_n^{-1}\left(1-\frac{\gamma_{mn}}{\Gamma_m}\right)
\frac{G^2}{\Gamma_n\sqrt{1+\ae}}\frac{1}{z_1-z_2}\times\nonumber\\
&&\times\left[\frac{z_1-\Gamma_\pm}
{z_1+iz}-\frac{z_2-\Gamma_\pm}{z_2+iz}\right].
\end{eqnarray}}
In the case of real $z_{1,2}$ the sign of $z_1-\Gamma_\pm$ and $z_2-\Gamma_\pm$ is
the same but depends on the sign of $\Gamma_0-\Gamma_\pm$. If $\Gamma_0>\Gamma_\pm$
then $z_{1,2}-\Gamma_\pm>0$; on the other hand, if $\Gamma_0<\Gamma_\pm$ then
$z_{1,2}-\Gamma_\pm<0$ (see (\ref{e3.4})). According to Fig.\ref{f4} of particular
interest is the case of strongly different $z_1$ and $z_2$
when $Re [F_\pm(z)]$ has the form of a broad dispersive contour
(the width $z_1$) with a sharp notch (or spike) in the
center (the width of $z_2\ll z_1$). The conditions that allow
for $z_1\gg z_2$ were analyzed above. We note that $z_2\ll z_1$
can be realized when ${\bf k}_\mu\cdot{\bf k}<0$.

If $z_{1,2}$ are complex, $Re [F_\pm(\Omega_\mu)]$ has the form

{\fs
\begin{eqnarray}
&&Re [F_\pm(\Omega_\mu)]=\frac{k_\mu}{k}\frac{z_0}{\Gamma_n}
\left(1-\frac{\gamma_{mn}}{\Gamma_m}\right)\frac{G^2}{\sqrt{1+\ae}}
\left\{\frac{1}{z_0^2+(z+\zeta)^2}\right.\\ \nonumber
&&\left. +\frac{1}{z_0^2+(z-\zeta)^2}-\frac{\Gamma_0-\Gamma_\pm}{\Gamma_0+\Gamma_\pm}
\frac{1}{\zeta}\left[\frac{z+\zeta}{z_0^2+(z+\zeta)^2}-
\frac{z-\zeta}{z_0^2+(z-\zeta)^2}\right]\right\}.\label{e3.15}
\end{eqnarray}}
In contrast to (\ref{e3.13}) the possibility to observe splitting
is determined now not only by the ratio $\zeta/z_0$ but also
by the magnitude and sign of the factor $(\Gamma_0-\Gamma_\pm)/(\Gamma_0+\Gamma_\pm)$.
From (\ref{e3.3}) for $\Gamma_0,\: \Gamma_\pm$ we can see that
$-1<s\equiv (\Gamma_0-\Gamma_\pm)/(\Gamma_0+\Gamma_\pm)<1$.
Figure \ref{f5} shows plots of

{\fs $$ F_\pm=\left[\frac{k_\mu}{k}\left(1-\frac{\gamma_{mn}}{\Gamma_m}\right)
\frac{G^2}{z_0\Gamma_n\sqrt{1+\ae}}\right]Re F_\pm $$} for the limiting values of the
factor $s$ and for $\Gamma_0=\Gamma_\pm$. According to Fig.\ref{f5}, a sharply
defined splitting effect can occur even with $\zeta=z_0$ which is less than the
possible limit of $\zeta\leq z_0\sqrt{3}$. Particularly significant is curve 3 in
Fig.\ref{f5} according to which the intensity is much lower in the center than in the
side maxima. Using (\ref{e3.15}) we can obtain for $\zeta=z_0$, ${\bf k}_\mu\cdot{\bf
k}<0$ and $k_\mu=k$:

{\fs
\begin{equation}\label{e3.16}
\frac{Re [F_-(0)]}{Re [F_-(\zeta)]}=\frac{5}{2}\frac{\Gamma_-}{\Gamma_0+2\Gamma_-}\approx
\frac{5}{2}\frac{\Gamma_{jm}}{\Gamma_{jn}+2\Gamma_{jm}+\Gamma\sqrt{1+\ae}}.
\end{equation}}
Consequently if $\Gamma_{jn}+\Gamma_B\gg\Gamma_{jm}$, the ratio (\ref{e3.16}) is much
smaller than unity. The condition $\Gamma_0\gg\Gamma_-$ corresponds to the value $s =
1$ and it can be satisfied for $\Gamma\sqrt{1+\ae}\gg\Gamma_{jm}$.

Comparison of $F_\pm(\Omega_\mu)$ and $f_\pm(\Omega_\mu)$. It is clear from
the preceding discussion that the frequency dependences
of $F_\pm$ and $f_\pm$ are similar in general and in some cases
one term can emphasize or, conversely, concentrate the
effects contributed by the other.

We now consider the properties of the sum $F_\pm$ and $f_\pm$
and determine the weight of each of the two terms. We
begin with the case of real roots $z_{1,2}$. In this case the
curves $Re [F_\pm(z)]$ and $Re [f_\pm(z)]$ are of the same type
throughout and we may limit the analysis to a single
point $z=0$ (maximum or minimum). From (\ref{e3.3}) and
(\ref{e3.4}) we find

{\fs
\begin{eqnarray}\label{e3.17}
&&Re [F_\pm(0)+f_\pm(0)]=\frac{k_\mu}{k}\frac{G^2\Gamma_\pm}{z_1z_2\sqrt{1+\ae}}
\left[\frac{2}{\Gamma_n}\left(1-\frac{\gamma_{mn}}{\Gamma_m}\right)+\right.\nonumber\\
&&\left. +\frac{1}{\Gamma_\pm}(1\mp\sqrt{1+\ae})\right].
\end{eqnarray}}
The first term in the brackets is associated with $f_\pm$ and
the second with $f_\pm$. The appearance of the factors $1/\Gamma_n$
and  $1/\Gamma_\pm$ is understandable: $1/\Gamma_n$ determines the time
of interaction of an atom at the n level with the field. An
analog of such an "accumulation time" for the interference term is the quantity
$1/\Gamma_\pm$.

In addition to the factor $1-\gamma_{mn}/\Gamma_m$, whose role was discussed above,
the relation between $F_\pm(0)$ and $f_\pm(0)$ depends on the relaxation constants,
field amplitude, direction of observation, and the ratio $k_\mu/k$. To observe NIE
even with $\gamma_{mn}\ll\Gamma_m$ the most convenient conditions obtain when $k_\mu
= -k$ and $\Gamma_{jm}\ll\Gamma_n$; furthermore its role increases with the rise in
field intensity. Conversely when $\bf k$ and ${\bf k}_\mu$ are parallel we can expect
an almost complete elimination of NIE because the inequality
$\Gamma_+\gg\Gamma_n[\sqrt{1+\ae}-1]/2$ can be assured by $\Gamma_{jm}\gg\Gamma_n$,
$\Gamma\gg\Gamma_n$, $\ae\ll 1$ and $k_\mu>k$. Therefore $Re [F_\pm]$ as well as $Re
[f_\pm]$ can be predominant depending on the values of the numerous variable
parameters.

If $z_{1,2}$ are complex the expression for $Re [F_\pm + f_\pm ]$
differs from (\ref{e3.15}) only by the substitution of factor $s$

{\fs
\begin{equation}\label{3.18}
c=\frac{\Gamma_0-\Gamma_\pm}{\Gamma_0+\Gamma_\pm}-\frac{\Gamma_n}{\Gamma_0+\Gamma_\pm}
\left(1-\frac{\gamma_{mn}}{\Gamma_m}\right)^{-1}[1\mp\sqrt{1+\ae}],
\end{equation}}
where the second term reflects the role of $Re [f_\pm]$. We
can show that the value of $c$ varies between +1 and -1
Therefore the total contour can be deformed within the
same limits as $Re [F_\pm]$ (see Fig.\ref{f5}).

We now consider $w_{nj}$ for the intermediate values of the angle $\theta$ between
$\bf k$ and ${\bf k}_\mu$. We denote the velocity component perpendicular to $\bf k$
by $\bf u$:

{\fs
\begin{equation}\label{e3.19}
\Omega_\mu '=\Omega_\mu-{\bf ku}\sin\theta-{\bf k}_\mu {\bf v}\cos\theta,\:
\Omega '=\Omega-{\bf kv}.
\end{equation}}
According to (\ref{e3.19}) the averaging with respect to $v$ leads
as before to (\ref{e3.3}), except that ${\bf k}_\mu$ must be replaced by
$k_\mu\cos\theta$ (apart from the common factor in $F_\pm$ and $f_\pm$)
and $\Omega_\mu$ by $\Omega_\mu-ku\sin\theta$. The subsequent averaging with
respect to $u$ can be carried out although only its result
is given here When the angles are small, $\theta\ll\Gamma_+/k\bar v$, $\Gamma_0/k\bar v$,
there is practically no variation of $w_{nj}$.

The same consideration applies to the angles $|\pi-\theta|\ll\Gamma_- /k {\bar v}$,
$\Gamma_0 / k {\bar v}$. When $|\theta|$
(or $|\pi-\theta|$) increases above the indicated values the spectral width of the functions
$F_\pm,\: f_\pm$ increases approximately as $k\bar v\sin\theta$ and reaches the
full Doppler width when $\theta\approx\pi/2$.
Since the integrated intensity of the correction to $w_{nj}$ due to strong field
does not depend on $\theta$, the amplitude of this correction is $k\bar v /\Gamma_0$
times lower than in
the above cases. All these phenomena are due to the fact that the strong field
represents a plane monochromatic wave and causes changes in the distribution of only
one velocity component Therefore the case of $\theta=0$ and the adjacent directions
of $k_\mu$ is the most interesting one.

Our analysis deals with the case where both fields
represent plane traveling waves. The experimenter may
find it convenient to use a strong field within the resonator of a
suitable gas laser$^{[4,11,12]}$. The strong field then
has the form of a standing wave and the pattern of events
is somewhat different. When the departure from resonance in the strong
field is greater than the width of
Bennett distribution ($|\Omega|>\Gamma_0,\: \Gamma_\pm$), one can regard the
two traveling waves as fully independent because they
interact with different groups of atoms. Therefore the
expression for $w_{nj}$ now contains, instead of $F_+(\Omega_\mu)+f_+(\Omega_\mu)$
or $F_-(\Omega_\mu)+f_-(\Omega_\mu)$, the sum of these terms

{\fs
\begin{equation}\label{e3.20}
F_+(\Omega_\mu)+f_+(\Omega_\mu)+F_-(\Omega_\mu)+f_-(\Omega_\mu).
\end{equation}}
All the singularities of the terms with indices + or -
are now at the distance $\pm\Omega k_\mu/k$ from the line center
(see definition of $z$ in (\ref{e3.3})) and they overlap. Thus all
that we said for the case of a strong field in the form of
a traveling wave remains valid for that of a standing
wave. At the same time different frequencies should
produce effects corresponding to "interference" and
"non-interference" directions.

On the other hand if the condition $|\Omega|>\Gamma$ does not
hold, the Bennett distributions stemming from two opposed waves
overlap and we have a different situation.
We can say that the additive property of nonlinear effects
due to opposed waves appears a priori in the first approximation
(with respect to $G^2$), i.e., (\ref{e3.20}) is valid if
$G^2$ is left in the expression for $F_\pm + f_\pm$ only in the form
of a common factor. The invariance of (\ref{e3.2}) in successive
approximations with respect to $G^2$' is due to the fact
that large fields generate a spatial inhomogeneity of the
medium (with a period of $\lambda/2$)$^{[13]}$. Consequently the
atomic probability amplitudes are subject to a form of
phase modulation and the atomic levels are split into a
number of sublevels larger than the two sublevels typical of
the traveling wave. The above modulation was
Investigated in $^{[15,18]}$ in the case of resonance fluorescence
and it was found that the emission spectrum
changed significantly.
\section{GENERATION IN THE PRESENCE OF EXTERNAL
FIELD} \setcounter{equation}{0} In Secs. 2 and 3 the fields that resonated with
transitions $n - j$, $g - n$, etc., were considered weak (Fig.\ref{f1}).
Experiments$^{[13]}$ showed that generation at these transitions was a convenient
method of studying NIE. Therefore we now consider generation at the $g - n$ transition
(since it was studied in$^{[13]}$) The unsaturated (with respect to $G_\mu$) gain at
the $g - n$ transition changes in an external field $G$ that is resonant with $m - n$
(see Sec. 3). To compute the generation power at $g - n$ we must know the saturation
function of the $g - n$ transition We can show that once the conditions

{\fs
\begin{equation}\label{e4.1}
|N_m -N_n|\frac{G^2}{\Gamma^2}\ll |N_g-N_n|,|N_m-N_n|\frac{G^4}{\Gamma^4}\ll |N_g-N_n|\frac
{G_\mu^2}{\Gamma^2}
\end{equation}}
are satisfied, saturation at the $g - n$ transition is the same as in the case of $G =
0$. Therefore the generation power is determined by the standard formula

{\fs
\begin{eqnarray}
\label{e4.2}&&\frac{\Gamma_n+\Gamma_g-\gamma_{ng}}{\Gamma_n\Gamma_g\Gamma_{ng}}G_\mu^2
=\left[1-\frac{\Delta N\exp\{\Omega_\mu^2/(k_\mu{\bar v})^2\}+\alpha}{N_g-N_n}\right]\times\nonumber\\
&&\times\left[1+\frac{\Gamma_{ng}^2}{\Gamma_{ng}^2+\Omega_\mu^2}\right]^{-1};\\
\label{e4.3}&&\alpha=\frac{k_\mu}{k}(N_m-N_n)G^2\left\{\frac{1-\gamma_{mn}/\Gamma_m}{\Gamma_n
\Gamma_0}\left[\frac{\Gamma_0^2}{\Gamma_0^2+(\Omega_\mu+k_\mu\Omega/k)^2}\right.\right.\nonumber\\
&&\left. +\frac{\Gamma_0^2}{\Gamma_0^2+(\Omega_\mu-k_\mu\Omega/k)^2}\right]+\frac{1}
{\Gamma+\Gamma_{gn}-\Gamma_{gm}}\times\nonumber\\
&&\times\left.\left[\frac{\Gamma_+}{\Gamma_+^2+(\Omega_\mu-k_\mu\Omega/k)^2}
-\frac{\Gamma_0}{\Gamma_0^2+(\Omega_\mu-k_\mu\Omega/k)^2}\right]\right\},\\
\label{e4.4}&&\Gamma_0=\Gamma_{gn}+\frac{k_\mu}{k}\Gamma,\Gamma_+=
\begin{cases}
\Gamma_{gm}k_\mu/k+(1-k_\mu/k)\Gamma_{gn},& k_\mu<k\\
\Gamma_{gm}+(k_\mu/k-1)\Gamma,& k_\mu>k
\end{cases}
\end{eqnarray}}
where $\Delta N$ is the threshold population difference for
$G = 0$ and $\Omega_\mu = 0$. In the absence of the external field
(\ref{e4.2}) determines the usual dependence of power on $\Omega_\mu$
with the "Lamb dip". The term $\alpha$ introduces an
additional spectral structure.

We consider the case when the role of atomic collisions is small,
so that $\Gamma+\Gamma_{gn}-\Gamma_{gm}=\Gamma_n$. A "spike"
or a "dip" (depending on the sign of $N_m - N_n$) then appears at
the frequency $\Omega_\mu=-\Omega k_\mu/k$

\begin{figure}
  \centering
\includegraphics[width=40mm]{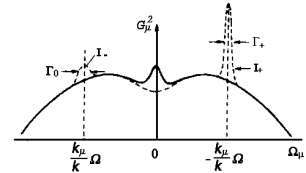}
\caption{Frequency dependence of generation power}\label{f6}
\end{figure}

{\fs
\begin{equation}\label{e4.5}
I_-=\frac{N_m-N_n}{N_n-N_g}\frac{k_\mu}{k}\frac{|G|^2}{\Gamma_n\Gamma_0}\left(
1-\frac{\gamma_{mn}}{\Gamma_m}\right)\frac{\Gamma_0^2}{\Gamma_0^2+(\Omega_\mu+k_\mu\Omega/k)^2}.
\end{equation}}
Another "spike" or "dip" appears at $\Omega_\mu=k_\mu\Omega/k$
(Fig.\ref{f6}).

{\fs
\begin{eqnarray}\label{e4.6}
&&I_+=\frac{N_m-N_n}{N_n-N_g}\frac{k_\mu}{k}\frac{|G|^2}{\Gamma_n\Gamma_0}
\left[\frac{\Gamma_0}{\Gamma_+}\frac{\Gamma_+^2}{\Gamma_+^2+(\Omega_\mu-k_\mu\Omega/k)^2}\right.
-\nonumber\\
&&\left. -\frac{\gamma_{mn}}{\Gamma_m}
\frac{\Gamma_0^2}{\Gamma_0^2+(\Omega_\mu-k_\mu\Omega/k)^2}\right];
\end{eqnarray}}
if $\Gamma_m,\: \Gamma_g\ll\Gamma_n$ and $|1-k_\mu/k|\ll 1$ then $\Gamma_+\ll\Gamma$
and $\Gamma_+\ll\Gamma_{gn}$ (see (ref{e4.4})). Consequently we see from (\ref{e4.5})
and (\ref{e4.6}) that in this case the "spikes" $I_-$ and $I_+$ differ sharply from
each other in width and height. The second term in (\ref{e4.6}) contributes
significantly only to the wings of the $I_+$, contour so that the width of this
"spike" is much smaller than the natural width at the $g - n$ transition. When
$\gamma_{mn}=\Gamma_m$ the "spike" $I_-$ vanishes and only the interference "spike"
$I_+$, remains with singularities in the wings (a "spike" in a "trough"). In the other
limiting case of $\Gamma_m\gg\Gamma_n,\: \Gamma_g$; $\Gamma_+\approx\Gamma_0$ both
spikes have the same width and vanish when $\gamma_{mn}/\Gamma_m\rightarrow 1$. When
$\Omega = 0$ and $\Gamma_+\ll\Gamma_{gn}$, the above singularities occur in the floor
of the Lamb "dip" as shown schematically in Fig.\ref{f6}.

Two generation peaks differing in width were observed in$^{[13]}$. A strong frequency
dependence of generation in the region $I_+$ can be utilized for effective output
power stabilization of generation frequency.

We consider the dependence of generating emission
frequency on the natural resonator frequency The generation frequency
is determined by the requirement that
the field phase shift in a double pass of the resonator be
a multiple of $2\pi$. The value of the refraction index
necessary to compute the phase can be found from

{\fs $$ n_0=1+2\pi NRe\{r_{ng}d_{ng}\}(E_\mu/4)^{-1}, $$} where $E_\mu$ is intensity
of the field resonating with the $n - g$ transition. If $|\Omega_\mu|\ll k_\mu{\bar
v}$ the generation frequency is determined from the equation

{\fs
\begin{eqnarray}
\label{e4.7}&&\Omega_r\equiv\omega_r-\omega_{gn}=\Omega_\mu+\frac{l}{l_r}\frac{\Delta\omega_r}
{2}\times\left\{\frac{2}{\sqrt{\pi}}\frac{N_g-N_n}{\Delta N}\frac{\Omega_\mu}{k{\bar
v}}
\right.\nonumber\\
&&\left. -\left[\frac{N_g-N_n}{\Delta N}-1\right]\frac{\Omega_\mu\Gamma_{ng}}{2\Gamma_{ng}^2-
\Omega_\mu^2}-\frac{k_\mu}{k}|G|^2\frac{N_m-N_n}{\Delta N}\Phi(\Omega_\mu)\right\},
\end{eqnarray}}
where $\omega_r$ is the natural frequency of the resonator and

\begin{figure}
  \centering
\includegraphics[width=40mm]{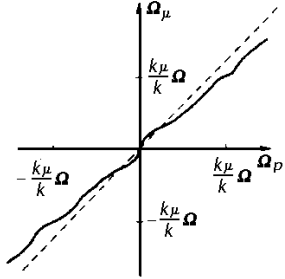}
\caption{Generation frequency as a function of resonator frequency.}\label{f7}
\end{figure}

{\fs
\begin{eqnarray*}
&&\Phi(\Omega_\mu)=\left(1-\frac{\gamma_{mn}}{\Gamma_m}\right)\frac{1}{\Gamma_n}\left[
\left(\Omega_\mu+\frac{k_\mu}{k}\Omega-\frac{\Gamma_0\Gamma_{ng}\Omega_\mu}{2\Gamma_{ng}^2
+\Omega_\mu^2}\right)\times\right.\\
&&\times\frac{1}{\Gamma_0^2+(\Omega_\mu+k_\mu\Omega /k)^2}+\\
&&\left. +\left(\Omega_\mu-\frac{k_\mu}{k}\Omega-\frac{\Gamma_0\Gamma_{ng}\Omega_\mu}
{2\Gamma_{ng}^2+\Omega_\mu^2}\right)\frac{1}{\Gamma_0^2+(\Omega_\mu-k_\mu\Omega /k)^2}\right]\\
&&+\frac{1}{\Gamma+\Gamma_{gn}-\Gamma_{gm}}\left[
\left(\Omega_\mu-\frac{k_\mu}{k}\Omega-\frac{\Gamma_+\Gamma_{gn}\Omega_\mu}{2\Gamma_{ng}^2
+\Omega_\mu^2}\right)\times\right.\\
&&\times\frac{1}{\Gamma_+^2+(\Omega_\mu-k_\mu\Omega /k)^2}-\\
&&\left. -\left(\Omega_\mu-\frac{k_\mu}{k}\Omega-\frac{\Gamma_0\Gamma_{gn}\Omega_\mu}
{2\Gamma_{gn}^2+\Omega_\mu^2}\right)\frac{1}{\Gamma_0^2+(\Omega_\mu-k_\mu\Omega /k)^2}\right].
\end{eqnarray*}}
The first term in the curved brackets of (\ref{e4.7}) describes the known phenomenon
of "pulling" the generation frequency by the natural resonator frequency towards the
center of the atomic line. The second describes a "repulsion" of the generation
frequency from the transition frequency towards the resonator frequency proportional
to the quantity $(N_g-N_n)/\Delta N -1$. On the curve of $\Omega_\mu$ as a function
of $\Omega_r$ (Fig.\ref{f7}) the first effect corresponds to the deviation of the
$\Omega_\mu$ asymptote from the straight line $\omega_r-\omega_{gn}=\Omega_\mu$ by an
angle of the order of $\Delta\omega_r/k_\mu{\bar v}$, and the second effect
corresponds to the singularity of the order of $\sqrt{2}\Gamma_{gn}$ near
$\Omega_\mu=0$.

We consider singularities occurring in the curve $\Omega_\mu$
in the region of frequencies $|\Omega_\mu\pm\Omega k_\mu/k|\lesssim\Gamma_{+,0}$ if
$\Gamma_{ng}\ll\Omega\ll k_\mu{\bar v}$. For a purely spontaneous relaxation
and $\gamma_{mn}\ll\Gamma_m$ we obtain from (\ref{e4.7})

{\fs
\begin{equation}\label{e4.9}
\Omega_{r^\pm}=\Omega_\mu-\frac{l_r}{l}\frac{\Delta\omega_r}{2}\frac{N_m-N_n}{\Delta
N}
\frac{k_\mu}{k}\frac{|G|^2}{\Gamma_n\Gamma_{+,0}}\frac{\Gamma_{+,0}(\Omega_\mu\mp\Omega
k_\mu/k)} {\Gamma_{+,0}^2+(\Omega_\mu\mp\Omega k_\mu/k)^2}.
\end{equation}}
The term proportional to $\Delta\omega_r/k_\mu{\bar v}$ has been dropped. It appears
from (\ref{e4.9}) that in the presence of an external field when
$\Omega_\mu=\pm\Omega k_\mu/k$ the dependence of generation frequency on the natural
resonator frequency increases when $N_m - N_n > 0$ and decreases when $N_m - N_n < 0$:

{\fs $$ \left(\frac{d\Omega_\mu}{d\Omega_\mu^\pm}\right)_{\Omega_\mu=\pm
k_\mu\Omega/k}=
\left[1-\frac{\Delta\omega_r}{2\Gamma_\pm}\frac{l}{l_r}\frac{N_m-N_n}{\Delta N}
\frac{k_\mu}{k}\frac{G^2}{\Gamma_n\Gamma_{+,0}}\right]. $$} In the latter case this
phenomenon can be used for passive stabilization of the generation frequency. The
lower the resonator $Q$ the greater this effect. If $\gamma_{mn}=\Gamma_m$ the
singularity at $\Omega_\mu=-\Omega k_\mu/k$ vanishes. At $\Omega =0$ all the
singularities in $\Omega_\mu$ as a function of $\Omega_r$ appear only when
$|\Omega_\mu|\lesssim\max\{\Gamma_{ng},\Gamma_0,\Gamma_+\}$. The dependence of
$\Omega_\mu$ on $\Omega_r$ can be cumbersome in this case. However if
$\Gamma_+\ll\Gamma_{ng},\Gamma_0$ the most pronounced is only the contribution from
$\Gamma_+$.

Translated by S. Kassel
97
\end{multicols}
\end{document}